\documentclass[nofootinbib,pra,twocolumn,superscriptaddress]{revtex4-2}
\usepackage{float,graphicx,multirow, amsmath,color,dsfont}
\usepackage{amsmath,bm}
\usepackage[colorlinks]{hyperref}
\usepackage[noabbrev]{cleveref}
\usepackage{enumerate}
\usepackage{amssymb,mathrsfs,dsfont}
\usepackage{amssymb,mathbbol}
\usepackage{amsmath}
\usepackage{amsfonts}
\usepackage{mathrsfs}
\usepackage{float}
\usepackage{xcolor,hyperref}
\definecolor{darkblue}{rgb}{0,0.0.1,0.3}
\definecolor{darkred}{rgb}{0.6,0.1,0}
\hypersetup{colorlinks,breaklinks,
	linkcolor=darkred,urlcolor=darkblue,
	anchorcolor=darkred,citecolor=
	darkred,pdfauthor=onoff, pdftitle= onoff}
\usepackage{tikz}
\usepackage{mathbbol}
\usepackage{array}
\usepackage{pifont}
\usepackage{accents}

\usepackage{braket}

\usepackage[normalem]{ulem} % For bibliography 
\newcommand{\ie}{\textit{i}.\textit{e}.}

\begin{document}

	\title{Comparing on-off detector and single photon detector in photon subtraction based continuous variable quantum teleportation}
	
	\author{Chandan Kumar}
	\email{chandan.quantum@gmail.com}
	\affiliation{Optics and Quantum Information Group, The Institute of Mathematical Sciences, CIT Campus, Taramani, Chennai 600113, India.}
	\affiliation{Homi Bhabha National Institute, Training School Complex, Anushakti Nagar, Mumbai 400085, India.}
	
	\author{Karunesh K. Mishra}
	\email{karunesh.mishra@eli-np.ro}
	\affiliation{Extreme Light Infrastructure - Nuclear Physics (ELI-NP),
		``Horia Hulubei'' National R\&D Institute for Physics and Nuclear Engineering (IFIN-HH),
		30 Reactorului Street, 077125 Bucharest-M\u{a}gurele, Romania}	
	\author{Sibasish Ghosh}
	\email{sibasish@imsc.res.in}
	\affiliation{Optics and Quantum Information Group, The Institute of Mathematical Sciences, CIT Campus, Taramani, Chennai 600113, India.}
	\affiliation{Homi Bhabha National Institute, Training School Complex, Anushakti Nagar, Mumbai 400085, India.}
	
	\begin{abstract}
		We consider here two distinct photon detectors namely, single photon detector  and on-off detector, to implement photon subtraction on a two-mode squeezed vacuum (TMSV) state. The two distinct  photon subtracted TMSV states generated are utilized individually as resource states in continuous variable quantum teleportation. Owing to the fact that the two generated states have different success probabilities (of photon subtraction) and fidelities (of quantum teleportation), we consider the product of the success probability  and fidelity enhancement as a figure of merit for the comparison of the two detectors.  The results show that the single photon detector should be preferred over the on-off detector for the maximization of the considered figure of merit. 
	\end{abstract}
	\maketitle
	%%%%%%%%%%%%%%%%%%%%%%%%%%%%%%%%%%
	\section{Introduction} 
	
	Photon subtraction (PS) have been shown to be useful in various quantum protocols such as squeezing and entanglement distillation~\cite{Ourjoumtsev,Takahashi2010,Lvovsky,Dirmeier:20,Kumar_2024}, quantum teleportation~\cite{tel2000,tel2009,Illuminati,catalysis15,catalysis17,wang2015,Ayan,tele-arxiv,better} and quantum metrology~\cite{gerryc-pra-2012,josab-2012,braun-pra-2014,josab-2016,pra-catalysis-2021,crs-ngtmsv-met,manali}.
	PS can be experimentally implemented via beam splitter and photon detector. We can choose either on-off detector or single photon detector (SPD)\cite{Braunstein_imperfect} to implement PS operation.
	Recently, a comparative study of the on-off detector and SPD-generated PS states have been conducted in the context of entanglement-based continuous variable (CV) quantum key distribution~\cite{Chen:23}. It turned out that the SPD-generated PS state renders higher transmission distances as compared to the on-off detector-generated PS state. 
	
	In CV quantum teleportation, usually  two mode squeezed vacuum (TMSV) state is used as resource states.  
	In order to improve the fidelity, PS operation is utilized. More specifically, 
	PS on both modes of TMSV state using SPD was performed and used as resource state for CV quantum teleportation~\cite{Opatrny, Cochrane}. Later, PS was performed using on-off detector  and used as resource state for CV quantum teleportation~\cite{Paris-pra-2003}. The former and the latter resource states will be called single-photon subtracted (SPS) TMSV state and inconclusive photon subtracted (IPS) TMSV state. 
	
	Two key quantities that differ significantly between these two states are the success probability of PS operation and the fidelity of teleportation. To properly evaluate the effectiveness of these states, it is essential to consider both the factors. Although the fidelity enhancements for both detectors are nearly the same, the on-off detector offers a higher success probability compared to the SPD for the generation of photon subtracted TMSV state~\cite{Paris-pra-2003}. This may initially suggest that the on-off detector is the preferable choice.
	
	To provide a more rigorous comparison of their performance, we consider a figure of merit defined as the product of the success probability and fidelity enhancement~\cite{tele-arxiv}. For ideal detectors, our results demonstrate that SPD achieves a higher value for this figure of merit. Consequently, the SPD emerges as the superior option over the on-off detector in the context of photon subtraction-based CV quantum teleportation.

	We also consider non-ideal detectors.
	The  SPD, utilizing  superconducting transition edge sensors, can achieve an impressive efficiency of up to 95 \%, while the on-off detector, operating in the visible regime, reaches only around  60 \%~\cite{Barbieri}.  Thus, even under realistic, non-ideal conditions, the SPD remains the superior choice compared to the on-off detector.
	
	The remaining sections of the paper are structured as follows.  In Sec.~\ref{back}, we discuss the  ideal and non-ideal photon detectors,   quantum teleportation protocol and PS operation. In Sec.~\ref{sec:detector}, we compare the fidelity of    SPS-TMSV and IPS-TMSV states    and in Sec.~\ref{success}, we compare the performance of two detectors using the considered figure of merit. Finally, we wrap up with a discussion in Sec.~\ref{sec:conc}, where we highlight the implications and future aspects of the current work.

	%%%%%%%%%%%%%%%%%%%%%%%%%%%%%%%%%%%%%%%%%%%
	
	\section{Background}\label{back}
	\subsection{Ideal and non-ideal photon detectors}
	\label{sec:POVM}
\begin{figure}[h!]
		\includegraphics[scale=1.1]{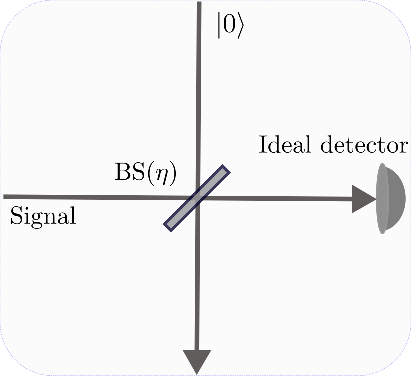}
		\caption{A non-ideal detector of efficiency $\eta$ can be modeled by a beam splitter of  transmissivity   $\eta$ followed by an ideal detector.   }
		\label{Imperfect}
	\end{figure}

	SPD can be represented by  two outcome projective measurement operators 
	given by $\{  |1\rangle\langle 1|,\,\,\mathbb{1}-|1\rangle\langle 1|\}$. Similarly,   an on-off detector is represented by the two-outcome projective measurement $\{{\Pi}_{\rm OFF},~ {\Pi}_{\rm ON}\}$, where
	\begin{eqnarray}
		\Pi_{\text{OFF}}=|0\rangle\langle 0|,
		\qquad
		\Pi_{\text{ON}}=\mathbb{1}-\Pi_{\text{OFF}}.
	\end{eqnarray} 
	A non-ideal detector can be modeled as a combination of a beam splitter of transmissivity   $\eta$ and an ideal detector, where $\eta$ represents the efficiency of the realistic detector.  The schematic is shown in Fig.~\ref{Imperfect}. We assume that the vacuum state enters the unused port of the beam splitter.

	In a non-ideal scenario, the projection operators  $|1\rangle\langle 1|$ of the ideal SPD gets replaced by
	\begin{eqnarray}
		\Pi_{1}(\eta)=\sum_{n=1}^{\infty} n \eta(1-\eta)^{n-1} |n\rangle\langle n|.  
	\end{eqnarray}
	Similarly, the ``ON"  element of the ideal on-off detector gets transformed into
	\begin{eqnarray}
		\Pi_\text{ON}(\eta)=\mathbb{1}-\sum_{n=0}^{\infty}(1-\eta)^{n} |n\rangle\langle n|.
	\end{eqnarray}

	\subsection{Continuous variable quantum teleportation protocol}
	\label{sec:qt}

\begin{figure}[h!]
		\includegraphics[scale=0.68]{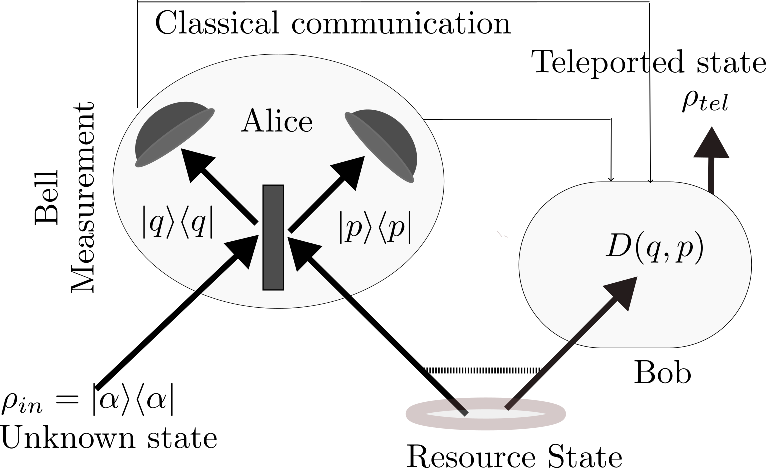}
		\caption{A schematic representation of the ideal Vaidman-Braunstein-Kimble  protocol for the teleportation of an unknown input coherent state. The shared entangled resource state can be TMSV, SPS-TMSV, or IPS-TMSV states. }
		\label{schematictele}
	\end{figure}

	Here we implement the Vaidman-Braunstein-Kimble  protocol for teleporting an unknown input quantum state between two distinct physical systems~\cite{Vaidman,bk-1998}. 
	An entangled resource is shared between Alice and Bob and Alice is given an unknown input quantum state for teleportation to Bob.  The density operator for the entangled resource state is denoted as $\rho$, whereas the unknown input state is denoted as $\rho_{\text{in}}$.   Alice combines her mode and the single-mode input state by employing a balanced beam splitter. Alice then performs ideal homodyne measurements of $\hat{q}$ and $\hat{p}$-quadratures on the two output modes, after which Alice communicates the results to Bob. Bob displaces his mode based on the results, and the resulting state is the teleported state denoted as $\rho_{\text{tel}}$. The schematic is shown in Fig.~\ref{schematictele}. 
	
	An appropriate choice for the fidelity of teleportation is 
	\begin{equation}
		F =~ {\rm Tr} \left[\sqrt{{\rho}_{\rm in}^{1/2} {\rho}_{\rm tel} {\rho}_{\rm in}^{1/2}}\right].
	\end{equation}
	When the unknown input states ${\rho}_{\rm in}$ are pure, the expression of fidelity reduces to $F =\text{Tr} [\rho_{\text{in}}\rho_{\text{tel}}]$.
	
	The TMSV state is considered as the  resource state for the teleportation of the input coherent state.
	The TMSV state is obtained by the action of two-mode squeezing operator on two single-mode vacuum state~\cite{tmsv1,tmsv2,Reid}:
	\begin{equation}
		|\psi\rangle=	\exp[r (\hat{a}_1^{\dagger} \hat{a}_2^{\dagger}-\hat{a}_1
		\hat{a}_2) ]|0\rangle_1|0\rangle_2,
	\end{equation}
	where $r$ is the (non-negative) squeezing parameter.
	The fidelity of quantum teleportation using the TMSV resource state turns out to be 
	\begin{equation}\label{tmsvf}
		F^{\text{TMSV}}=\frac{\lambda +1}{2}, \quad \lambda=\tanh{r}.
	\end{equation}

	Teleporting an input coherent state can be executed with a maximum fidelity of $1/2$ without the need for a shared entangled state~\cite{Braunstein-jmo-2000,Braunstein-pra-2001}. 
	Therefore, the success of quantum teleportation  is determined by the fidelity magnitude exceeding the classical limit of $1/2$. From Eq.~(\ref{tmsvf}), we see that quantum teleportation is achieved for $r>0$. Further,  to accomplish perfect teleportation with unit fidelity, we need a TMSV state with infinite squeezing, which is impractical.
	
	\subsection{Photon subtraction using on-off detector and single photon detector}\label{ps}
	
	\begin{figure}[h!]
		\includegraphics[scale=1]{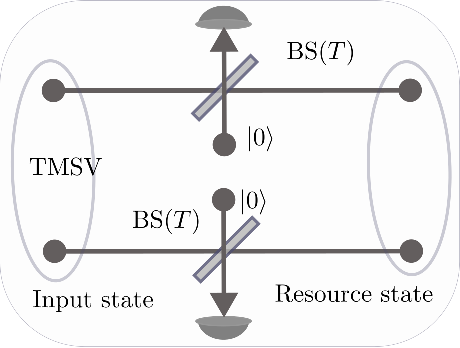}
		\caption{Implementation of PS operation on the input TMSV state. PS operation on both the modes of the TMSV is implemented via a beam splitter and photon detector. The detector  can be either on-off detector or single photon detector.   }
		\label{figsub}
	\end{figure}
	
	We intend to generate  photon subtracted TMSV state using on-off detector and SPD and compare their utility in CV quantum teleportation. 
	The experimental setup for the generation of  photon subtracted TMSV
	states is shown in Fig.~\ref{figsub}. Each mode of the TMSV state is combined with
	vacumm state $|0\rangle$  using a  beam splitter of transmissivity $T$. On the output mode corresponding to the vacuum state, we can either use the on-off detector and SPD. When the ``on" element of the on-off detector clicks, IPS-TMSV state is generated. Similarly, the detector of a single photon via SPD heralds the generation of SPS-TMSV.

	\section{Comparison of two detectors} 
	We first compare the fidelity of teleporting input coherent state using SPS-TMSV and IPS-TMSV resource states and then move on to compare the product of success probability and fidelity enhancement for the two detectors. 
	\subsection{Fidelity using ideal and non-ideal detectors}
	\label{sec:detector}
\begin{figure}[h!]
		\includegraphics[scale=1]{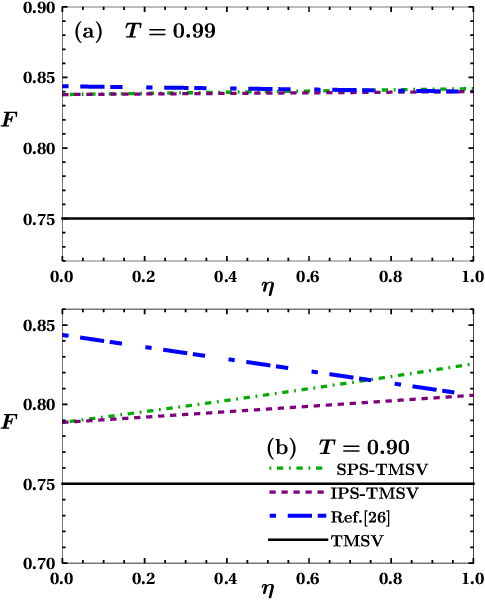}
		\caption{ Non-ideal detector: Fidelity as a function of detector efficiency $\eta$ for different transmissivities. We have set  $\lambda=0.5$. }
		\label{fidelity_tau_v01}
	\end{figure}

	We have provided  the analytical expression of the fidelity of teleporting input coherent state using photon subtracted TMSV state, when ideal detectors are employed, in Table~\ref{table1}. Additionally, we have also provided the  analytical expression of the success probability of the two distinct PS operations.
	\begin{widetext}
		
		\begin{table}[H] 
			\centering
			\caption{\label{table1} The fidelity of   teleporting   input coherent state using the IPS-TMSV and SPS-TMSV resource states and the success probability of generating these resource states.}
			\renewcommand{\arraystretch}{2.5}
			\begin{tabular}{ |c|c|c| }
				\hline \hline
				& Fidelity & Success probability \\ \hline \hline
				IPS-TMSV & $F^{\text{ON-OFF}} = \displaystyle\frac{(\lambda +1)(\lambda T +1) (2-(2-\lambda ) \lambda T ) \left(1-\lambda ^2T \right)}{2 (\lambda  (1-T )+1) \left(\lambda ^2T +1\right) (2-\lambda T  (\lambda  (1-\tau )+2))}$ & $P^{\text{ON-OFF}} = \displaystyle\frac{\lambda^2(1-T)^2\left(\lambda^2T+1\right)}{\left(1-\lambda ^2T\right)\left(1-\lambda^2T^2\right)}$ \\ \hline 
				%--------------------------------------------------------------
				SPS-TMSV & $F^{\text{SPD}} = \displaystyle\frac{(\lambda T +1)^3 (2-\lambda T  (2-\lambda T ))}{4 \left(\lambda ^2T ^2+1\right)}$ & $P^{\text{SPD}} = \displaystyle\frac{\lambda^2\left(1-\lambda^2\right)(1-T)^2\left(\lambda^2T ^2+1\right)}{\left(1-\lambda^2T^2\right)^3}$ \\ \hline \hline
			\end{tabular}
		\end{table}

		\begin{figure}[htbp]
			\includegraphics[scale=1]{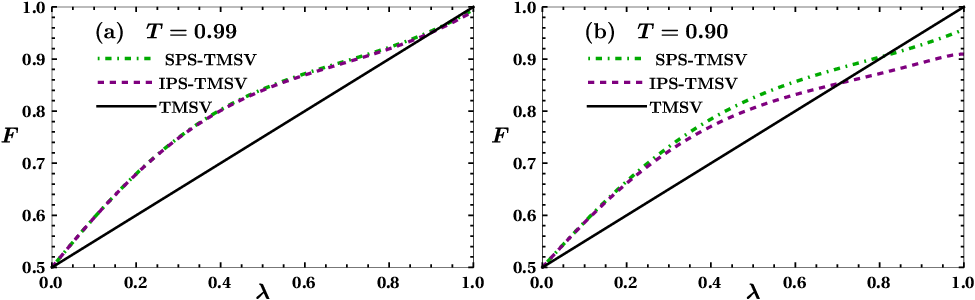}
			\caption{Ideal detector ($\eta=1$): Fidelity as a function of squeezing parameter $\lambda$ for different transmissivities.  }
			\label{fidelity_lambda_v01}
		\end{figure}
		
		We have shown the fidelity as a function of squeezing parameter $\lambda$ for different transmissivities ($T=0.99$ and  0.90) in Fig.~\ref{fidelity_lambda_v01}. At  $T=0.9$, SPS-TMSV state yields better fidelity than IPS-TMSV. As $T$ approaches unity, the fidelity difference between SPS-TMSV and IPS-TMSV state reduces. In fact, the fidelity is equal for both the states   in the unit transmissivity limit:
		\begin{equation}\label{maxf}
			\lim\limits_{T \to 1} F^{\text{ON-OFF}}=\lim\limits_{T \to 1}F^{\text{SPD}}=\frac{(\lambda +1)^3 (2-(2-\lambda ) \lambda )}{4 \left(\lambda ^2+1\right)} .
		\end{equation}
		We also note that fidelity for both states is maximized in the unit transmissivity limit.

		We now turn to non-ideal detectors.  The success probability of generating SPS-TMSV and IPS-TMSV states  can be determined by substituting the transmissivity $T$ with $T_{\text{eff.}}=1-\eta(1-T)$ in the success probability of the ideal detector.
		The expression of fidelity of SPS-TMSV  generated using non-ideal detector can be obtained as
		\begin{equation}
			\label{Eq:Fid_non_ideal_SPD}
			F^{\text{SPD}}= \frac{(\lambda -\eta  \lambda  (1-T)+1)^3 \left(\lambda ^2 \left(2 \eta ^2 (1-T)^2-2 \eta  (2-T) (1-T)-(2-T) T+2\right)-2 \lambda  T+2\right)}{4 ((1-\eta) \lambda  (1-T)+1)^3 \left((\lambda +\eta  \lambda  (1-T))^2+1\right)}.
		\end{equation}
		Similarly, the expression of fidelity of IPS-TMSV  generated using non-ideal detector can be obtained as
		\begin{equation}
			\label{Eq:Fid_non_ideal_ONOFF}
			F^{\text{ON-OFF}}= \frac{(\lambda+1) (1+\lambda -(1-T)\eta\lambda) \left(1-(1-\eta(1-T))\lambda ^2\right) (2-\lambda  (-2 \lambda -\lambda  (2-T) ((1-\eta ) (-T)-\eta )+2 T))}
			%----------------------------------------
			{2(\lambda(1-T)+1)((1-\eta)\lambda(1-T)+1) \left(2-\lambda ^2 (1-T) (2-\eta  (2-T))-2 \lambda T\right) \left(\lambda^2(1-\eta(1-T))+1\right)} .
		\end{equation}
		
	\end{widetext}
	We have presented the plot that illustrates the relationship between fidelity and detector efficiency $\eta$ in Fig.~\ref{fidelity_tau_v01} for different transmissivities  with the squeezing being constant.

	For the SPS-TMSV state, the fidelity tends to decrease when the detector efficiency decreases. The same trend can be observed with the IPS-TMSV state. However, it is worth noting that the fidelity of the SPS-TMSV state decreases at a higher rate compared to the IPS-TMSV state. 
	
	We have also shown the graph of IPS-TMSV state using the analytical expression provided in Ref.~\cite{Paris-pra-2003}. In contrast with our result,  the fidelity of the IPS-TMSV state increases with a decrease in the detector efficiency $\eta$. This occurs because the authors simply replaced the transmissivity $T$ with $T_{\text{eff.}}=1-\eta(1-T)$ in the ideal detector fidelity to obtain the fidelity for the non-ideal detector. While such a substitution leads to the correct result for the success probability, it does not lead to the correct expression of the teleportation fidelity.

	\begin{table*}
		\centering
		\caption{\label{table2}
			Maximum value of the product $\mathcal{R}$  for teleportation of the input coherent state using  photon subtracted TMSV  state for ideal  and non-ideal detectors. We have also shown the corresponding squeezing, transmissivity, fidelity enhancement, and success probability.   }
		\renewcommand{\arraystretch}{1.8}
		\begin{tabular}
			{ |p{3cm}|>{\centering\arraybackslash} p{1.5cm}|>{\centering\arraybackslash}p{1cm}|>{\centering\arraybackslash}p{1cm}|p{1cm}||p{1cm}||p{1cm}|}
			\hline \hline
			& $10^{4} \times \mathcal{R}_\text{max}$&  $\lambda$ & $T$ &   ${\Delta}F$ & $10 \times P$ \\ \hline \hline
			ON-OFF ($\eta=1$) & 3.9 & 0.49&0.84& 0.033 &0.12\\ \hline
			ON-OFF ($\eta=0.60$) & 1.1& 0.47&0.85& 0.032 &0.04\\ \hline  \hline
			SPD ($\eta=1$) &9.5  & 0.56 &0.77 & 0.037&0.26\\ \hline 
			SPD ($\eta=0.95$) &7.6 & 0.55 &0.77 & 0.036&0.21\\ \hline 
			\hline
		\end{tabular}
	\end{table*} 
	
	\subsection{Contour plot of success probability and fidelity enhancement}\label{success}
	In the preceding sections, we studied the fidelity of the SPS-TMSV and IPS-TMSV states for fixed values of transmissivity and/or squeezing.

	Since the PS operation is probabilistic in nature, we need to consider the success probability to compare the two distinct types of photon-subtracted TMSV states. 
	
\begin{figure}[h!]
		\includegraphics[scale=1]{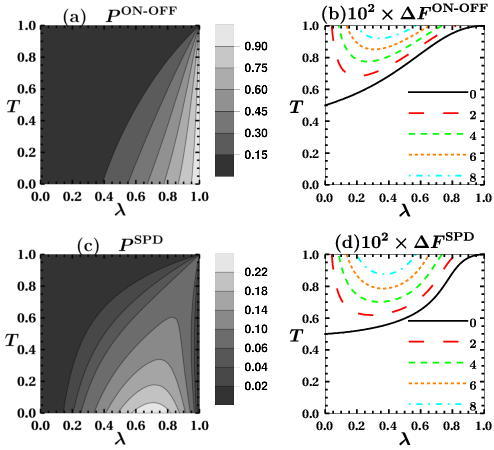}
		\caption{ Ideal detector ($\eta=1$): First column shows the success probability of PS operations on the TMSV state as a function of the transmissivity ($T$) and the squeezing parameter ($\lambda$). 
			Second column shows  the fidelity enhancement ($\Delta{F}$) achieved in the teleportation of input coherent states utilizing the  photon subtracted TMSV resource states. 
		}
		\label{prob_fid}
	\end{figure}

	In order to find the squeezing and transmissivity values where implementing the PS  operation enhances the fidelity, we define a figure of merit as the difference in teleportation fidelities between the photon subtracted TMSV states and the TMSV state as
	\begin{equation}
		\Delta F^{\text{ON-OFF/SPD}} = F^{\text{ON-OFF/SPD}}-F^{\text{TMSV}} .
	\end{equation}
	Our interest is the region in the $(\lambda,T)$ space, where  $\Delta F^{\text{ON-OFF/SPD}}$ is positive. 
	Along with the enhancement in fidelity, we also consider the success probability.  In Fig~\ref{prob_fid},
	we have shown success probability and  curves of fixed $\Delta F^{\text{ON-OFF/SPD}}$ in the $(\lambda,T)$ space. The success probability of PS using an on-off detector is observed to be greater than that of PS utilizing  SPD.    
	On the other hand, the magnitude of the enhancement of fidelity is nearly equal for the two different detectors.

	Since the teleportation fidelity for the TMSV state is independent of transmissivity,  the fidelity enhancement is also maximized in the unit transmissivity limit along with the fidelity (see Eq.~(\ref{maxf})). On the other hand, we can see from Table~\ref{table1} that  the success probability ($P^\text{ON-OFF}$ and $P^\text{SPD}$) approaches zero in the unit transmissivity limit. Such a scenario is not practically feasible.
	
	\subsection{Product of success probability and fidelity enhancement}
\begin{figure}[h!]
		\includegraphics[scale=1]{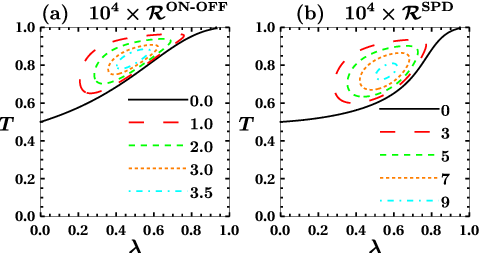}
		\caption{ Ideal detector: The product of success probability and fidelity enhancement ($\mathcal{R}$) as a function of the transmissivity ($T$) and the squeezing parameter ($\lambda$).}
		\label{prob_times_dif_fid}
	\end{figure}

	To get the optimal scenario, the enhancement in fidelity can be traded off against the success probability. For quantitative analysis, we define the product of success probability and fidelity enhancement as a figure of merit:
	\begin{equation}
		\mathcal{R} =P \times \Delta F = P \times (F^{\text{ON-OFF/SPD}}-F^{\text{TMSV}} ).
	\end{equation} 
	We show curves of fixed values of $\mathcal{R}$ in Fig.~\ref{prob_times_dif_fid} for the two distinct detectors.  We see that the maximum product value achieved by the SPS-TMSV state is higher as compared to the corresponding value for the IPS-TMSV state.  In Table~\ref{table2}, we provide the magnitude of various parameters that yield the maximum value of the product $\mathcal{R}$.  
	
	Hence, performing the PS operation using an SPD detector is preferable to using an on-off detector. Given that SPD detectors ($\eta=0.95$) exhibit significantly higher efficiency compared to on-off detectors ($\eta=0.60$), this further strengthens the case for utilizing SPD. Additionally, we have presented the maximum value of the product $\mathcal{R}$ for non-ideal detectors in Table~\ref{table2}.

	\subsection{  Enhancement in fidelity vs. average photon number}
	\label{avg_photon_number}
	In this subsection, we intend to study the relation between enhancement  in fidelity and average photon number.
	For two mode system, the average photon number can be written as 
	\begin{equation}
		\hat{N} = \hat{a}_1^\dagger\hat{a}_1+\hat{a}_2^\dagger \hat{a}_2
	\end{equation}
	For the TMSV and the SPS-TMSV states, the average photon number is  
	\begin{equation}
		\langle \hat{N} \rangle^{\text{TMSV}}=  \frac{2 \lambda ^2}{1-\lambda ^2},\quad   \langle\hat{N}\rangle^{\text{ SPD}} =    \frac{4 \lambda ^2 T^2 \left(\lambda ^2 T^2+2\right)}{1-\lambda ^4 T^4 }.
	\end{equation}
	We define enhancement of average photon number as,
	\begin{equation}
		\Delta\langle\hat{N}\rangle^{\text{ SPD}}=\langle\hat{N}\rangle^{\text{ SPD}}-\langle \hat{N} \rangle^{\text{TMSV}} .
	\end{equation}
	
	\begin{figure}[htbp]
		\includegraphics[scale=1]{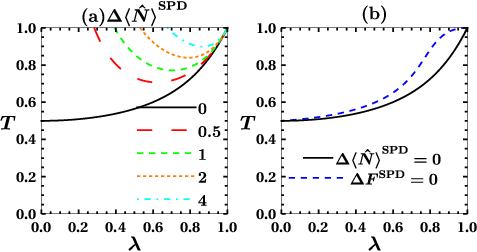}
		\caption{Ideal detector: (a) The enhancement of the average photon number ($\Delta\langle \hat{N} \rangle ^{\text{ SPD}}$)     as function of $\lambda$ and $T$ for the SPS-TMSV state. (b) In the enclosed region between the two curves,  the average photon number increases but the fidelity does not increase.}
		\label{avg_and_fid}
	\end{figure}

	In Fig.~\ref{avg_and_fid}(a), we show the enhancement of the average photon number  
	of the SPS-TMSV state  as function of $\lambda$ and $T$. In Fig.~\ref{avg_and_fid}(b), we show the curves $\Delta\langle \hat{N} \rangle ^{\text{ SPD}} =0 $ and  $\Delta F ^{\text{ SPD}} =0 $ on the $(\lambda, \tau)$ space. In the area bounded by the two curves, there is an increment in the average photon number increases however the fidelity decreases.

\begin{figure}[H]
		\includegraphics[scale=1]{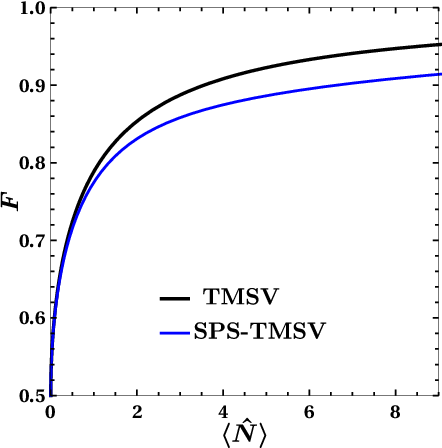}
		\caption{Ideal detector: Teleportation fidelity as a function of average   photon number.}
		\label{fvsmean1}
	\end{figure}

	We analyze the fidelity as a function of the average photon number for the TMSV  and SPS-TMSV states in Fig.~\ref{fvsmean1}. For the SPS-TMSV state, we have taken unit transmissivity, \ie, the state is effectively 
	$\propto \hat{a}_1 \hat{a}_2 |\text{TMSV}\rangle$. We see that the fidelity increases with average photon number. However, 
	for a fixed average  photon number, the TMSV state performs better that the SPS-TMSV state.

	\section{Conclusion}
	\label{sec:conc}
	We conducted a comparative analysis of the efficacy of on-off detectors versus SPD detectors in performing PS operations, where the resulting photon-subtracted states are employed as a resource for CV quantum teleportation. While the fidelity enhancement achieved by both detectors is nearly identical, the on-off detector shows a higher success probability than the SPD. This might initially suggest that the on-off detector could surpass the SPD in terms of performance.

	To quantitatively compare the two detectors, we consider the product of the success probability and the fidelity enhancement in teleportation as the key figure of merit. Upon careful examination, however, it becomes clear that SPD maximizes this figure of merit. Therefore, despite the higher success probability of the on-off detector, SPD is the superior choice for CV quantum teleportation.

	Our result is in line with a recent comparative study of on-off detectors and SPD-generated photon-subtracted TMSV states in the context of entanglement-based CV quantum key distribution~\cite{Chen:23}.
	This work will allow experimentalists to choose appropriate detectors along with other parameters for performance optimization. 
	We can compare the impact of two detectors in other quantum protocols such as entanglement distillation~\cite{Distillation}, dense coding scheme~\cite{dense}, quantum illumination~\cite{illumination} and quantum imaging~\cite{Imaging}.

%%%%%%%%%%%%%%%%%%%%%%%%%%%%%%%%%%%%%%%%%%%%%	 
%%%%%%%%%%%%%%%%%%%%%%%%%%%%%%%%%%%%%%%%%%%%%
	
%\bibliography{references}

\begin{thebibliography}{38}%
		\makeatletter
		\providecommand \@ifxundefined [1]{%
			\@ifx{#1\undefined}
		}%
		\providecommand \@ifnum [1]{%
			\ifnum #1\expandafter \@firstoftwo
			\else \expandafter \@secondoftwo
			\fi
		}%
		\providecommand \@ifx [1]{%
			\ifx #1\expandafter \@firstoftwo
			\else \expandafter \@secondoftwo
			\fi
		}%
		\providecommand \natexlab [1]{#1}%
		\providecommand \enquote  [1]{``#1''}%
		\providecommand \bibnamefont  [1]{#1}%
		\providecommand \bibfnamefont [1]{#1}%
		\providecommand \citenamefont [1]{#1}%
		\providecommand \href@noop [0]{\@secondoftwo}%
		\providecommand \href [0]{\begingroup \@sanitize@url \@href}%
		\providecommand \@href[1]{\@@startlink{#1}\@@href}%
		\providecommand \@@href[1]{\endgroup#1\@@endlink}%
		\providecommand \@sanitize@url [0]{\catcode `\\12\catcode `\$12\catcode
			`\&12\catcode `\#12\catcode `\^12\catcode `\_12\catcode `\%12\relax}%
		\providecommand \@@startlink[1]{}%
		\providecommand \@@endlink[0]{}%
		\providecommand \url  [0]{\begingroup\@sanitize@url \@url }%
		\providecommand \@url [1]{\endgroup\@href {#1}{\urlprefix }}%
		\providecommand \urlprefix  [0]{URL }%
		\providecommand \Eprint [0]{\href }%
		\providecommand \doibase [0]{https://doi.org/}%
		\providecommand \selectlanguage [0]{\@gobble}%
		\providecommand \bibinfo  [0]{\@secondoftwo}%
		\providecommand \bibfield  [0]{\@secondoftwo}%
		\providecommand \translation [1]{[#1]}%
		\providecommand \BibitemOpen [0]{}%
		\providecommand \bibitemStop [0]{}%
		\providecommand \bibitemNoStop [0]{.\EOS\space}%
		\providecommand \EOS [0]{\spacefactor3000\relax}%
		\providecommand \BibitemShut  [1]{\csname bibitem#1\endcsname}%
		\let\auto@bib@innerbib\@empty
		%</preamble>
		\bibitem [{\citenamefont {Ourjoumtsev}\ \emph {et~al.}(2007)\citenamefont
			{Ourjoumtsev}, \citenamefont {Dantan}, \citenamefont {Tualle-Brouri},\ and\
			\citenamefont {Grangier}}]{Ourjoumtsev}%
		\BibitemOpen
		\bibfield  {author} {\bibinfo {author} {\bibfnamefont {A.}~\bibnamefont
				{Ourjoumtsev}}, \bibinfo {author} {\bibfnamefont {A.}~\bibnamefont {Dantan}},
			\bibinfo {author} {\bibfnamefont {R.}~\bibnamefont {Tualle-Brouri}},\ and\
			\bibinfo {author} {\bibfnamefont {P.}~\bibnamefont {Grangier}},\ }\bibfield
		{title} {\bibinfo {title} {Increasing entanglement between gaussian states by
				coherent photon subtraction},\ }\href
		{https://doi.org/10.1103/PhysRevLett.98.030502} {\bibfield  {journal}
			{\bibinfo  {journal} {Phys. Rev. Lett.}\ }\textbf {\bibinfo {volume} {98}},\
			\bibinfo {pages} {030502} (\bibinfo {year} {2007})}\BibitemShut {NoStop}%
		\bibitem [{\citenamefont {Takahashi}\ \emph {et~al.}(2010)\citenamefont
			{Takahashi}, \citenamefont {Neergaard-Nielsen}, \citenamefont {Takeuchi},
			\citenamefont {Takeoka}, \citenamefont {Hayasaka}, \citenamefont {Furusawa},\
			and\ \citenamefont {Sasaki}}]{Takahashi2010}%
		\BibitemOpen
		\bibfield  {author} {\bibinfo {author} {\bibfnamefont {H.}~\bibnamefont
				{Takahashi}}, \bibinfo {author} {\bibfnamefont {J.~S.}\ \bibnamefont
				{Neergaard-Nielsen}}, \bibinfo {author} {\bibfnamefont {M.}~\bibnamefont
				{Takeuchi}}, \bibinfo {author} {\bibfnamefont {M.}~\bibnamefont {Takeoka}},
			\bibinfo {author} {\bibfnamefont {K.}~\bibnamefont {Hayasaka}}, \bibinfo
			{author} {\bibfnamefont {A.}~\bibnamefont {Furusawa}},\ and\ \bibinfo
			{author} {\bibfnamefont {M.}~\bibnamefont {Sasaki}},\ }\bibfield  {title}
		{\bibinfo {title} {Entanglement distillation from gaussian input states},\
		}\href {https://doi.org/10.1038/nphoton.2010.1} {\bibfield  {journal}
			{\bibinfo  {journal} {Nature Photonics}\ }\textbf {\bibinfo {volume} {4}},\
			\bibinfo {pages} {178} (\bibinfo {year} {2010})}\BibitemShut {NoStop}%
		\bibitem [{\citenamefont {Kurochkin}\ \emph {et~al.}(2014)\citenamefont
			{Kurochkin}, \citenamefont {Prasad},\ and\ \citenamefont
			{Lvovsky}}]{Lvovsky}%
		\BibitemOpen
		\bibfield  {author} {\bibinfo {author} {\bibfnamefont {Y.}~\bibnamefont
				{Kurochkin}}, \bibinfo {author} {\bibfnamefont {A.~S.}\ \bibnamefont
				{Prasad}},\ and\ \bibinfo {author} {\bibfnamefont {A.~I.}\ \bibnamefont
				{Lvovsky}},\ }\bibfield  {title} {\bibinfo {title} {Distillation of the
				two-mode squeezed state},\ }\href
		{https://doi.org/10.1103/PhysRevLett.112.070402} {\bibfield  {journal}
			{\bibinfo  {journal} {Phys. Rev. Lett.}\ }\textbf {\bibinfo {volume} {112}},\
			\bibinfo {pages} {070402} (\bibinfo {year} {2014})}\BibitemShut {NoStop}%
		\bibitem [{\citenamefont {Dirmeier}\ \emph {et~al.}(2020)\citenamefont
			{Dirmeier}, \citenamefont {Tiedau}, \citenamefont {Khan}, \citenamefont
			{Ansari}, \citenamefont {M\"{u}ller}, \citenamefont {Silberhorn},
			\citenamefont {Marquardt},\ and\ \citenamefont {Leuchs}}]{Dirmeier:20}%
		\BibitemOpen
		\bibfield  {author} {\bibinfo {author} {\bibfnamefont {T.}~\bibnamefont
				{Dirmeier}}, \bibinfo {author} {\bibfnamefont {J.}~\bibnamefont {Tiedau}},
			\bibinfo {author} {\bibfnamefont {I.}~\bibnamefont {Khan}}, \bibinfo {author}
			{\bibfnamefont {V.}~\bibnamefont {Ansari}}, \bibinfo {author} {\bibfnamefont
				{C.~R.}\ \bibnamefont {M\"{u}ller}}, \bibinfo {author} {\bibfnamefont
				{C.}~\bibnamefont {Silberhorn}}, \bibinfo {author} {\bibfnamefont
				{C.}~\bibnamefont {Marquardt}},\ and\ \bibinfo {author} {\bibfnamefont
				{G.}~\bibnamefont {Leuchs}},\ }\bibfield  {title} {\bibinfo {title}
			{Distillation of squeezing using an engineered pulsed parametric
				down-conversion source},\ }\href {https://doi.org/10.1364/OE.402178}
		{\bibfield  {journal} {\bibinfo  {journal} {Opt. Express}\ }\textbf {\bibinfo
				{volume} {28}},\ \bibinfo {pages} {30784} (\bibinfo {year}
			{2020})}\BibitemShut {NoStop}%
		\bibitem [{\citenamefont {Kumar}(2024)}]{Kumar_2024}%
		\BibitemOpen
		\bibfield  {author} {\bibinfo {author} {\bibfnamefont {C.}~\bibnamefont
				{Kumar}},\ }\bibfield  {title} {\bibinfo {title} {Advantage of probabilistic
				non-gaussian operations in the distillation of single mode squeezed vacuum
				state},\ }\href {https://doi.org/10.1088/1402-4896/ad6ae4} {\bibfield
			{journal} {\bibinfo  {journal} {Physica Scripta}\ }\textbf {\bibinfo {volume}
				{99}},\ \bibinfo {pages} {095112} (\bibinfo {year} {2024})}\BibitemShut
		{NoStop}%
		\bibitem [{\citenamefont {Opatrn\'y}\ \emph
			{et~al.}(2000{\natexlab{a}})\citenamefont {Opatrn\'y}, \citenamefont
			{Kurizki},\ and\ \citenamefont {Welsch}}]{tel2000}%
		\BibitemOpen
		\bibfield  {author} {\bibinfo {author} {\bibfnamefont {T.}~\bibnamefont
				{Opatrn\'y}}, \bibinfo {author} {\bibfnamefont {G.}~\bibnamefont {Kurizki}},\
			and\ \bibinfo {author} {\bibfnamefont {D.-G.}\ \bibnamefont {Welsch}},\
		}\bibfield  {title} {\bibinfo {title} {Improvement on teleportation of
				continuous variables by photon subtraction via conditional measurement},\
		}\href {https://doi.org/10.1103/PhysRevA.61.032302} {\bibfield  {journal}
			{\bibinfo  {journal} {Phys. Rev. A}\ }\textbf {\bibinfo {volume} {61}},\
			\bibinfo {pages} {032302} (\bibinfo {year} {2000}{\natexlab{a}})}\BibitemShut
		{NoStop}%
		\bibitem [{\citenamefont {Yang}\ and\ \citenamefont {Li}(2009)}]{tel2009}%
		\BibitemOpen
		\bibfield  {author} {\bibinfo {author} {\bibfnamefont {Y.}~\bibnamefont
				{Yang}}\ and\ \bibinfo {author} {\bibfnamefont {F.-L.}\ \bibnamefont {Li}},\
		}\bibfield  {title} {\bibinfo {title} {Entanglement properties of
				non-gaussian resources generated via photon subtraction and addition and
				continuous-variable quantum-teleportation improvement},\ }\href
		{https://doi.org/10.1103/PhysRevA.80.022315} {\bibfield  {journal} {\bibinfo
				{journal} {Phys. Rev. A}\ }\textbf {\bibinfo {volume} {80}},\ \bibinfo
			{pages} {022315} (\bibinfo {year} {2009})}\BibitemShut {NoStop}%
		\bibitem [{\citenamefont {Dell'Anno}\ \emph {et~al.}(2007)\citenamefont
			{Dell'Anno}, \citenamefont {De~Siena}, \citenamefont {Albano},\ and\
			\citenamefont {Illuminati}}]{Illuminati}%
		\BibitemOpen
		\bibfield  {author} {\bibinfo {author} {\bibfnamefont {F.}~\bibnamefont
				{Dell'Anno}}, \bibinfo {author} {\bibfnamefont {S.}~\bibnamefont {De~Siena}},
			\bibinfo {author} {\bibfnamefont {L.}~\bibnamefont {Albano}},\ and\ \bibinfo
			{author} {\bibfnamefont {F.}~\bibnamefont {Illuminati}},\ }\bibfield  {title}
		{\bibinfo {title} {Continuous-variable quantum teleportation with
				non-gaussian resources},\ }\href {https://doi.org/10.1103/PhysRevA.76.022301}
		{\bibfield  {journal} {\bibinfo  {journal} {Phys. Rev. A}\ }\textbf {\bibinfo
				{volume} {76}},\ \bibinfo {pages} {022301} (\bibinfo {year}
			{2007})}\BibitemShut {NoStop}%
		\bibitem [{\citenamefont {Xu}(2015)}]{catalysis15}%
		\BibitemOpen
		\bibfield  {author} {\bibinfo {author} {\bibfnamefont {X.-x.}\ \bibnamefont
				{Xu}},\ }\bibfield  {title} {\bibinfo {title} {Enhancing quantum entanglement
				and quantum teleportation for two-mode squeezed vacuum state by local
				quantum-optical catalysis},\ }\href
		{https://doi.org/10.1103/PhysRevA.92.012318} {\bibfield  {journal} {\bibinfo
				{journal} {Phys. Rev. A}\ }\textbf {\bibinfo {volume} {92}},\ \bibinfo
			{pages} {012318} (\bibinfo {year} {2015})}\BibitemShut {NoStop}%
		\bibitem [{\citenamefont {Hu}\ \emph {et~al.}(2017)\citenamefont {Hu},
			\citenamefont {Liao},\ and\ \citenamefont {Zubairy}}]{catalysis17}%
		\BibitemOpen
		\bibfield  {author} {\bibinfo {author} {\bibfnamefont {L.}~\bibnamefont
				{Hu}}, \bibinfo {author} {\bibfnamefont {Z.}~\bibnamefont {Liao}},\ and\
			\bibinfo {author} {\bibfnamefont {M.~S.}\ \bibnamefont {Zubairy}},\
		}\bibfield  {title} {\bibinfo {title} {Continuous-variable entanglement via
				multiphoton catalysis},\ }\href {https://doi.org/10.1103/PhysRevA.95.012310}
		{\bibfield  {journal} {\bibinfo  {journal} {Phys. Rev. A}\ }\textbf {\bibinfo
				{volume} {95}},\ \bibinfo {pages} {012310} (\bibinfo {year}
			{2017})}\BibitemShut {NoStop}%
		\bibitem [{\citenamefont {Wang}\ \emph {et~al.}(2015)\citenamefont {Wang},
			\citenamefont {Hou}, \citenamefont {Chen},\ and\ \citenamefont
			{Xu}}]{wang2015}%
		\BibitemOpen
		\bibfield  {author} {\bibinfo {author} {\bibfnamefont {S.}~\bibnamefont
				{Wang}}, \bibinfo {author} {\bibfnamefont {L.-L.}\ \bibnamefont {Hou}},
			\bibinfo {author} {\bibfnamefont {X.-F.}\ \bibnamefont {Chen}},\ and\
			\bibinfo {author} {\bibfnamefont {X.-F.}\ \bibnamefont {Xu}},\ }\bibfield
		{title} {\bibinfo {title} {Continuous-variable quantum teleportation with
				non-gaussian entangled states generated via multiple-photon subtraction and
				addition},\ }\href {https://doi.org/10.1103/PhysRevA.91.063832} {\bibfield
			{journal} {\bibinfo  {journal} {Phys. Rev. A}\ }\textbf {\bibinfo {volume}
				{91}},\ \bibinfo {pages} {063832} (\bibinfo {year} {2015})}\BibitemShut
		{NoStop}%
		\bibitem [{\citenamefont {Patra}\ \emph {et~al.}(2022)\citenamefont {Patra},
			\citenamefont {Gupta}, \citenamefont {Roy},\ and\ \citenamefont
			{Sen(De)}}]{Ayan}%
		\BibitemOpen
		\bibfield  {author} {\bibinfo {author} {\bibfnamefont {A.}~\bibnamefont
				{Patra}}, \bibinfo {author} {\bibfnamefont {R.}~\bibnamefont {Gupta}},
			\bibinfo {author} {\bibfnamefont {S.}~\bibnamefont {Roy}},\ and\ \bibinfo
			{author} {\bibfnamefont {A.}~\bibnamefont {Sen(De)}},\ }\bibfield  {title}
		{\bibinfo {title} {Significance of fidelity deviation in continuous-variable
				teleportation},\ }\href {https://doi.org/10.1103/PhysRevA.106.022433}
		{\bibfield  {journal} {\bibinfo  {journal} {Phys. Rev. A}\ }\textbf {\bibinfo
				{volume} {106}},\ \bibinfo {pages} {022433} (\bibinfo {year}
			{2022})}\BibitemShut {NoStop}%
		\bibitem [{\citenamefont {Kumar}\ and\ \citenamefont
			{Arora}(2023)}]{tele-arxiv}%
		\BibitemOpen
		\bibfield  {author} {\bibinfo {author} {\bibfnamefont {C.}~\bibnamefont
				{Kumar}}\ and\ \bibinfo {author} {\bibfnamefont {S.}~\bibnamefont {Arora}},\
		}\bibfield  {title} {\bibinfo {title} {Success probability and performance
				optimization in non-gaussian continuous-variable quantum teleportation},\
		}\href {https://doi.org/10.1103/PhysRevA.107.012418} {\bibfield  {journal}
			{\bibinfo  {journal} {Phys. Rev. A}\ }\textbf {\bibinfo {volume} {107}},\
			\bibinfo {pages} {012418} (\bibinfo {year} {2023})}\BibitemShut {NoStop}%
		\bibitem [{\citenamefont {Kumar}\ \emph {et~al.}(2024)\citenamefont {Kumar},
			\citenamefont {Sharma},\ and\ \citenamefont {Arora}}]{better}%
		\BibitemOpen
		\bibfield  {author} {\bibinfo {author} {\bibfnamefont {C.}~\bibnamefont
				{Kumar}}, \bibinfo {author} {\bibfnamefont {M.}~\bibnamefont {Sharma}},\ and\
			\bibinfo {author} {\bibfnamefont {S.}~\bibnamefont {Arora}},\ }\bibfield
		{title} {\bibinfo {title} {Continuous variable quantum teleportation in a
				dissipative environment: Comparison of non-gaussian operations before and
				after noisy channel},\ }\href
		{https://doi.org/https://doi.org/10.1002/qute.202300344} {\bibfield
			{journal} {\bibinfo  {journal} {Advanced Quantum Technologies}\ }\textbf
			{\bibinfo {volume} {7}},\ \bibinfo {pages} {2300344} (\bibinfo {year}
			{2024})}\BibitemShut {NoStop}%
		\bibitem [{\citenamefont {Birrittella}\ \emph {et~al.}(2012)\citenamefont
			{Birrittella}, \citenamefont {Mimih},\ and\ \citenamefont
			{Gerry}}]{gerryc-pra-2012}%
		\BibitemOpen
		\bibfield  {author} {\bibinfo {author} {\bibfnamefont {R.}~\bibnamefont
				{Birrittella}}, \bibinfo {author} {\bibfnamefont {J.}~\bibnamefont {Mimih}},\
			and\ \bibinfo {author} {\bibfnamefont {C.~C.}\ \bibnamefont {Gerry}},\
		}\bibfield  {title} {\bibinfo {title} {Multiphoton quantum interference at a
				beam splitter and the approach to heisenberg-limited interferometry},\ }\href
		{https://doi.org/10.1103/PhysRevA.86.063828} {\bibfield  {journal} {\bibinfo
				{journal} {Phys. Rev. A}\ }\textbf {\bibinfo {volume} {86}},\ \bibinfo
			{pages} {063828} (\bibinfo {year} {2012})}\BibitemShut {NoStop}%
		\bibitem [{\citenamefont {Carranza}\ and\ \citenamefont
			{Gerry}(2012)}]{josab-2012}%
		\BibitemOpen
		\bibfield  {author} {\bibinfo {author} {\bibfnamefont {R.}~\bibnamefont
				{Carranza}}\ and\ \bibinfo {author} {\bibfnamefont {C.~C.}\ \bibnamefont
				{Gerry}},\ }\bibfield  {title} {\bibinfo {title} {Photon-subtracted two-mode
				squeezed vacuum states and applications to quantum optical interferometry},\
		}\href {https://doi.org/10.1364/JOSAB.29.002581} {\bibfield  {journal}
			{\bibinfo  {journal} {J. Opt. Soc. Am. B}\ }\textbf {\bibinfo {volume}
				{29}},\ \bibinfo {pages} {2581} (\bibinfo {year} {2012})}\BibitemShut
		{NoStop}%
		\bibitem [{\citenamefont {Braun}\ \emph {et~al.}(2014)\citenamefont {Braun},
			\citenamefont {Jian}, \citenamefont {Pinel},\ and\ \citenamefont
			{Treps}}]{braun-pra-2014}%
		\BibitemOpen
		\bibfield  {author} {\bibinfo {author} {\bibfnamefont {D.}~\bibnamefont
				{Braun}}, \bibinfo {author} {\bibfnamefont {P.}~\bibnamefont {Jian}},
			\bibinfo {author} {\bibfnamefont {O.}~\bibnamefont {Pinel}},\ and\ \bibinfo
			{author} {\bibfnamefont {N.}~\bibnamefont {Treps}},\ }\bibfield  {title}
		{\bibinfo {title} {Precision measurements with photon-subtracted or
				photon-added gaussian states},\ }\href
		{https://doi.org/10.1103/PhysRevA.90.013821} {\bibfield  {journal} {\bibinfo
				{journal} {Phys. Rev. A}\ }\textbf {\bibinfo {volume} {90}},\ \bibinfo
			{pages} {013821} (\bibinfo {year} {2014})}\BibitemShut {NoStop}%
		\bibitem [{\citenamefont {Ouyang}\ \emph {et~al.}(2016)\citenamefont {Ouyang},
			\citenamefont {Wang},\ and\ \citenamefont {Zhang}}]{josab-2016}%
		\BibitemOpen
		\bibfield  {author} {\bibinfo {author} {\bibfnamefont {Y.}~\bibnamefont
				{Ouyang}}, \bibinfo {author} {\bibfnamefont {S.}~\bibnamefont {Wang}},\ and\
			\bibinfo {author} {\bibfnamefont {L.}~\bibnamefont {Zhang}},\ }\bibfield
		{title} {\bibinfo {title} {Quantum optical interferometry via the
				photon-added two-mode squeezed vacuum states},\ }\href
		{https://doi.org/10.1364/JOSAB.33.001373} {\bibfield  {journal} {\bibinfo
				{journal} {J. Opt. Soc. Am. B}\ }\textbf {\bibinfo {volume} {33}},\ \bibinfo
			{pages} {1373} (\bibinfo {year} {2016})}\BibitemShut {NoStop}%
		\bibitem [{\citenamefont {Zhang}\ \emph {et~al.}(2021)\citenamefont {Zhang},
			\citenamefont {Ye}, \citenamefont {Wei}, \citenamefont {Xia}, \citenamefont
			{Chang}, \citenamefont {Liao},\ and\ \citenamefont
			{Hu}}]{pra-catalysis-2021}%
		\BibitemOpen
		\bibfield  {author} {\bibinfo {author} {\bibfnamefont {H.}~\bibnamefont
				{Zhang}}, \bibinfo {author} {\bibfnamefont {W.}~\bibnamefont {Ye}}, \bibinfo
			{author} {\bibfnamefont {C.}~\bibnamefont {Wei}}, \bibinfo {author}
			{\bibfnamefont {Y.}~\bibnamefont {Xia}}, \bibinfo {author} {\bibfnamefont
				{S.}~\bibnamefont {Chang}}, \bibinfo {author} {\bibfnamefont
				{Z.}~\bibnamefont {Liao}},\ and\ \bibinfo {author} {\bibfnamefont
				{L.}~\bibnamefont {Hu}},\ }\bibfield  {title} {\bibinfo {title} {Improved
				phase sensitivity in a quantum optical interferometer based on multiphoton
				catalytic two-mode squeezed vacuum states},\ }\href
		{https://doi.org/10.1103/PhysRevA.103.013705} {\bibfield  {journal} {\bibinfo
				{journal} {Phys. Rev. A}\ }\textbf {\bibinfo {volume} {103}},\ \bibinfo
			{pages} {013705} (\bibinfo {year} {2021})}\BibitemShut {NoStop}%
		\bibitem [{\citenamefont {Kumar}\ \emph {et~al.}(2022)\citenamefont {Kumar},
			\citenamefont {Rishabh},\ and\ \citenamefont {Arora}}]{crs-ngtmsv-met}%
		\BibitemOpen
		\bibfield  {author} {\bibinfo {author} {\bibfnamefont {C.}~\bibnamefont
				{Kumar}}, \bibinfo {author} {\bibnamefont {Rishabh}},\ and\ \bibinfo {author}
			{\bibfnamefont {S.}~\bibnamefont {Arora}},\ }\bibfield  {title} {\bibinfo
			{title} {Realistic non-gaussian-operation scheme in parity-detection-based
				mach-zehnder quantum interferometry},\ }\href
		{https://doi.org/10.1103/PhysRevA.105.052437} {\bibfield  {journal} {\bibinfo
				{journal} {Phys. Rev. A}\ }\textbf {\bibinfo {volume} {105}},\ \bibinfo
			{pages} {052437} (\bibinfo {year} {2022})}\BibitemShut {NoStop}%
		\bibitem [{\citenamefont {Verma}\ \emph {et~al.}(2023)\citenamefont {Verma},
			\citenamefont {Kumar}, \citenamefont {Mishra},\ and\ \citenamefont
			{Panigrahi}}]{manali}%
		\BibitemOpen
		\bibfield  {author} {\bibinfo {author} {\bibfnamefont {M.}~\bibnamefont
				{Verma}}, \bibinfo {author} {\bibfnamefont {C.}~\bibnamefont {Kumar}},
			\bibinfo {author} {\bibfnamefont {K.~K.}\ \bibnamefont {Mishra}},\ and\
			\bibinfo {author} {\bibfnamefont {P.~K.}\ \bibnamefont {Panigrahi}},\
		}\bibfield  {title} {\bibinfo {title} {Optimal non-gaussian operations in
				difference-intensity detection and parity detection-based mach-zehnder
				interferometer},\ }\href@noop {} {\bibfield  {journal} {\bibinfo  {journal}
				{arXiv preprint arXiv:2312.10774}\ } (\bibinfo {year} {2023})}\BibitemShut
		{NoStop}%
		\bibitem [{\citenamefont {Nemoto}\ and\ \citenamefont
			{Braunstein}(2002)}]{Braunstein_imperfect}%
		\BibitemOpen
		\bibfield  {author} {\bibinfo {author} {\bibfnamefont {K.}~\bibnamefont
				{Nemoto}}\ and\ \bibinfo {author} {\bibfnamefont {S.~L.}\ \bibnamefont
				{Braunstein}},\ }\bibfield  {title} {\bibinfo {title} {Equivalent efficiency
				of a simulated photon-number detector},\ }\href
		{https://doi.org/10.1103/PhysRevA.66.032306} {\bibfield  {journal} {\bibinfo
				{journal} {Phys. Rev. A}\ }\textbf {\bibinfo {volume} {66}},\ \bibinfo
			{pages} {032306} (\bibinfo {year} {2002})}\BibitemShut {NoStop}%
		\bibitem [{\citenamefont {Chen}\ \emph {et~al.}(2023)\citenamefont {Chen},
			\citenamefont {Jia}, \citenamefont {Zhao}, \citenamefont {Zhou},
			\citenamefont {Liu},\ and\ \citenamefont {Hu}}]{Chen:23}%
		\BibitemOpen
		\bibfield  {author} {\bibinfo {author} {\bibfnamefont {X.}~\bibnamefont
				{Chen}}, \bibinfo {author} {\bibfnamefont {F.}~\bibnamefont {Jia}}, \bibinfo
			{author} {\bibfnamefont {T.}~\bibnamefont {Zhao}}, \bibinfo {author}
			{\bibfnamefont {N.}~\bibnamefont {Zhou}}, \bibinfo {author} {\bibfnamefont
				{S.}~\bibnamefont {Liu}},\ and\ \bibinfo {author} {\bibfnamefont
				{L.}~\bibnamefont {Hu}},\ }\bibfield  {title} {\bibinfo {title}
			{Continuous-variable quantum key distribution based on non-gaussian
				operations with on-off detection},\ }\href
		{https://doi.org/10.1364/OE.493328} {\bibfield  {journal} {\bibinfo
				{journal} {Opt. Express}\ }\textbf {\bibinfo {volume} {31}},\ \bibinfo
			{pages} {32935} (\bibinfo {year} {2023})}\BibitemShut {NoStop}%
		\bibitem [{\citenamefont {Opatrn\'y}\ \emph
			{et~al.}(2000{\natexlab{b}})\citenamefont {Opatrn\'y}, \citenamefont
			{Kurizki},\ and\ \citenamefont {Welsch}}]{Opatrny}%
		\BibitemOpen
		\bibfield  {author} {\bibinfo {author} {\bibfnamefont {T.}~\bibnamefont
				{Opatrn\'y}}, \bibinfo {author} {\bibfnamefont {G.}~\bibnamefont {Kurizki}},\
			and\ \bibinfo {author} {\bibfnamefont {D.-G.}\ \bibnamefont {Welsch}},\
		}\bibfield  {title} {\bibinfo {title} {Improvement on teleportation of
				continuous variables by photon subtraction via conditional measurement},\
		}\href {https://doi.org/10.1103/PhysRevA.61.032302} {\bibfield  {journal}
			{\bibinfo  {journal} {Phys. Rev. A}\ }\textbf {\bibinfo {volume} {61}},\
			\bibinfo {pages} {032302} (\bibinfo {year} {2000}{\natexlab{b}})}\BibitemShut
		{NoStop}%
		\bibitem [{\citenamefont {Cochrane}\ \emph {et~al.}(2002)\citenamefont
			{Cochrane}, \citenamefont {Ralph},\ and\ \citenamefont {Milburn}}]{Cochrane}%
		\BibitemOpen
		\bibfield  {author} {\bibinfo {author} {\bibfnamefont {P.~T.}\ \bibnamefont
				{Cochrane}}, \bibinfo {author} {\bibfnamefont {T.~C.}\ \bibnamefont
				{Ralph}},\ and\ \bibinfo {author} {\bibfnamefont {G.~J.}\ \bibnamefont
				{Milburn}},\ }\bibfield  {title} {\bibinfo {title} {Teleportation improvement
				by conditional measurements on the two-mode squeezed vacuum},\ }\href
		{https://doi.org/10.1103/PhysRevA.65.062306} {\bibfield  {journal} {\bibinfo
				{journal} {Phys. Rev. A}\ }\textbf {\bibinfo {volume} {65}},\ \bibinfo
			{pages} {062306} (\bibinfo {year} {2002})}\BibitemShut {NoStop}%
		\bibitem [{\citenamefont {Olivares}\ \emph {et~al.}(2003)\citenamefont
			{Olivares}, \citenamefont {Paris},\ and\ \citenamefont
			{Bonifacio}}]{Paris-pra-2003}%
		\BibitemOpen
		\bibfield  {author} {\bibinfo {author} {\bibfnamefont {S.}~\bibnamefont
				{Olivares}}, \bibinfo {author} {\bibfnamefont {M.~G.~A.}\ \bibnamefont
				{Paris}},\ and\ \bibinfo {author} {\bibfnamefont {R.}~\bibnamefont
				{Bonifacio}},\ }\bibfield  {title} {\bibinfo {title} {Teleportation
				improvement by inconclusive photon subtraction},\ }\href
		{https://doi.org/10.1103/PhysRevA.67.032314} {\bibfield  {journal} {\bibinfo
				{journal} {Phys. Rev. A}\ }\textbf {\bibinfo {volume} {67}},\ \bibinfo
			{pages} {032314} (\bibinfo {year} {2003})}\BibitemShut {NoStop}%
		\bibitem [{\citenamefont {Barbieri}(2022)}]{Barbieri}%
		\BibitemOpen
		\bibfield  {author} {\bibinfo {author} {\bibfnamefont {M.}~\bibnamefont
				{Barbieri}},\ }\bibfield  {title} {\bibinfo {title} {Optical quantum
				metrology},\ }\href {https://doi.org/10.1103/PRXQuantum.3.010202} {\bibfield
			{journal} {\bibinfo  {journal} {PRX Quantum}\ }\textbf {\bibinfo {volume}
				{3}},\ \bibinfo {pages} {010202} (\bibinfo {year} {2022})}\BibitemShut
		{NoStop}%
		\bibitem [{\citenamefont {Vaidman}(1994)}]{Vaidman}%
		\BibitemOpen
		\bibfield  {author} {\bibinfo {author} {\bibfnamefont {L.}~\bibnamefont
				{Vaidman}},\ }\bibfield  {title} {\bibinfo {title} {Teleportation of quantum
				states},\ }\href {https://doi.org/10.1103/PhysRevA.49.1473} {\bibfield
			{journal} {\bibinfo  {journal} {Phys. Rev. A}\ }\textbf {\bibinfo {volume}
				{49}},\ \bibinfo {pages} {1473} (\bibinfo {year} {1994})}\BibitemShut
		{NoStop}%
		\bibitem [{\citenamefont {Braunstein}\ and\ \citenamefont
			{Kimble}(1998)}]{bk-1998}%
		\BibitemOpen
		\bibfield  {author} {\bibinfo {author} {\bibfnamefont {S.~L.}\ \bibnamefont
				{Braunstein}}\ and\ \bibinfo {author} {\bibfnamefont {H.~J.}\ \bibnamefont
				{Kimble}},\ }\bibfield  {title} {\bibinfo {title} {Teleportation of
				continuous quantum variables},\ }\href
		{https://doi.org/10.1103/PhysRevLett.80.869} {\bibfield  {journal} {\bibinfo
				{journal} {Phys. Rev. Lett.}\ }\textbf {\bibinfo {volume} {80}},\ \bibinfo
			{pages} {869} (\bibinfo {year} {1998})}\BibitemShut {NoStop}%
		\bibitem [{\citenamefont {Caves}\ and\ \citenamefont
			{Schumaker}(1985)}]{tmsv1}%
		\BibitemOpen
		\bibfield  {author} {\bibinfo {author} {\bibfnamefont {C.~M.}\ \bibnamefont
				{Caves}}\ and\ \bibinfo {author} {\bibfnamefont {B.~L.}\ \bibnamefont
				{Schumaker}},\ }\bibfield  {title} {\bibinfo {title} {New formalism for
				two-photon quantum optics. i. quadrature phases and squeezed states},\ }\href
		{https://doi.org/10.1103/PhysRevA.31.3068} {\bibfield  {journal} {\bibinfo
				{journal} {Phys. Rev. A}\ }\textbf {\bibinfo {volume} {31}},\ \bibinfo
			{pages} {3068} (\bibinfo {year} {1985})}\BibitemShut {NoStop}%
		\bibitem [{\citenamefont {Schumaker}\ and\ \citenamefont
			{Caves}(1985)}]{tmsv2}%
		\BibitemOpen
		\bibfield  {author} {\bibinfo {author} {\bibfnamefont {B.~L.}\ \bibnamefont
				{Schumaker}}\ and\ \bibinfo {author} {\bibfnamefont {C.~M.}\ \bibnamefont
				{Caves}},\ }\bibfield  {title} {\bibinfo {title} {New formalism for
				two-photon quantum optics. ii. mathematical foundation and compact
				notation},\ }\href {https://doi.org/10.1103/PhysRevA.31.3093} {\bibfield
			{journal} {\bibinfo  {journal} {Phys. Rev. A}\ }\textbf {\bibinfo {volume}
				{31}},\ \bibinfo {pages} {3093} (\bibinfo {year} {1985})}\BibitemShut
		{NoStop}%
		\bibitem [{\citenamefont {Reid}(1989)}]{Reid}%
		\BibitemOpen
		\bibfield  {author} {\bibinfo {author} {\bibfnamefont {M.~D.}\ \bibnamefont
				{Reid}},\ }\bibfield  {title} {\bibinfo {title} {Demonstration of the
				einstein-podolsky-rosen paradox using nondegenerate parametric
				amplification},\ }\href {https://doi.org/10.1103/PhysRevA.40.913} {\bibfield
			{journal} {\bibinfo  {journal} {Phys. Rev. A}\ }\textbf {\bibinfo {volume}
				{40}},\ \bibinfo {pages} {913} (\bibinfo {year} {1989})}\BibitemShut
		{NoStop}%
		\bibitem [{\citenamefont {Braunstein}\ \emph {et~al.}(2000)\citenamefont
			{Braunstein}, \citenamefont {Fuchs},\ and\ \citenamefont
			{Kimble}}]{Braunstein-jmo-2000}%
		\BibitemOpen
		\bibfield  {author} {\bibinfo {author} {\bibfnamefont {S.~L.}\ \bibnamefont
				{Braunstein}}, \bibinfo {author} {\bibfnamefont {C.~A.}\ \bibnamefont
				{Fuchs}},\ and\ \bibinfo {author} {\bibfnamefont {H.~J.}\ \bibnamefont
				{Kimble}},\ }\bibfield  {title} {\bibinfo {title} {Criteria for
				continuous-variable quantum teleportation},\ }\href
		{https://doi.org/10.1080/09500340008244041} {\bibfield  {journal} {\bibinfo
				{journal} {Journal of Modern Optics}\ }\textbf {\bibinfo {volume} {47}},\
			\bibinfo {pages} {267} (\bibinfo {year} {2000})}\BibitemShut {NoStop}%
		\bibitem [{\citenamefont {Braunstein}\ \emph {et~al.}(2001)\citenamefont
			{Braunstein}, \citenamefont {Fuchs}, \citenamefont {Kimble},\ and\
			\citenamefont {van Loock}}]{Braunstein-pra-2001}%
		\BibitemOpen
		\bibfield  {author} {\bibinfo {author} {\bibfnamefont {S.~L.}\ \bibnamefont
				{Braunstein}}, \bibinfo {author} {\bibfnamefont {C.~A.}\ \bibnamefont
				{Fuchs}}, \bibinfo {author} {\bibfnamefont {H.~J.}\ \bibnamefont {Kimble}},\
			and\ \bibinfo {author} {\bibfnamefont {P.}~\bibnamefont {van Loock}},\
		}\bibfield  {title} {\bibinfo {title} {Quantum versus classical domains for
				teleportation with continuous variables},\ }\href
		{https://doi.org/10.1103/PhysRevA.64.022321} {\bibfield  {journal} {\bibinfo
				{journal} {Phys. Rev. A}\ }\textbf {\bibinfo {volume} {64}},\ \bibinfo
			{pages} {022321} (\bibinfo {year} {2001})}\BibitemShut {NoStop}%
		\bibitem [{\citenamefont {Zhang}\ and\ \citenamefont {van
				Loock}(2010)}]{Distillation}%
		\BibitemOpen
		\bibfield  {author} {\bibinfo {author} {\bibfnamefont {S.~L.}\ \bibnamefont
				{Zhang}}\ and\ \bibinfo {author} {\bibfnamefont {P.}~\bibnamefont {van
					Loock}},\ }\bibfield  {title} {\bibinfo {title} {Distillation of mixed-state
				continuous-variable entanglement by photon subtraction},\ }\href
		{https://doi.org/10.1103/PhysRevA.82.062316} {\bibfield  {journal} {\bibinfo
				{journal} {Phys. Rev. A}\ }\textbf {\bibinfo {volume} {82}},\ \bibinfo
			{pages} {062316} (\bibinfo {year} {2010})}\BibitemShut {NoStop}%
		\bibitem [{\citenamefont {Kitagawa}\ \emph {et~al.}(2006)\citenamefont
			{Kitagawa}, \citenamefont {Takeoka}, \citenamefont {Sasaki},\ and\
			\citenamefont {Chefles}}]{dense}%
		\BibitemOpen
		\bibfield  {author} {\bibinfo {author} {\bibfnamefont {A.}~\bibnamefont
				{Kitagawa}}, \bibinfo {author} {\bibfnamefont {M.}~\bibnamefont {Takeoka}},
			\bibinfo {author} {\bibfnamefont {M.}~\bibnamefont {Sasaki}},\ and\ \bibinfo
			{author} {\bibfnamefont {A.}~\bibnamefont {Chefles}},\ }\bibfield  {title}
		{\bibinfo {title} {Entanglement evaluation of non-gaussian states generated
				by photon subtraction from squeezed states},\ }\href
		{https://doi.org/10.1103/PhysRevA.73.042310} {\bibfield  {journal} {\bibinfo
				{journal} {Phys. Rev. A}\ }\textbf {\bibinfo {volume} {73}},\ \bibinfo
			{pages} {042310} (\bibinfo {year} {2006})}\BibitemShut {NoStop}%
		\bibitem [{\citenamefont {Zhang}\ \emph {et~al.}(2014)\citenamefont {Zhang},
			\citenamefont {Guo}, \citenamefont {Bao}, \citenamefont {Shi}, \citenamefont
			{Jin}, \citenamefont {Zou},\ and\ \citenamefont {Guo}}]{illumination}%
		\BibitemOpen
		\bibfield  {author} {\bibinfo {author} {\bibfnamefont {S.}~\bibnamefont
				{Zhang}}, \bibinfo {author} {\bibfnamefont {J.}~\bibnamefont {Guo}}, \bibinfo
			{author} {\bibfnamefont {W.}~\bibnamefont {Bao}}, \bibinfo {author}
			{\bibfnamefont {J.}~\bibnamefont {Shi}}, \bibinfo {author} {\bibfnamefont
				{C.}~\bibnamefont {Jin}}, \bibinfo {author} {\bibfnamefont {X.}~\bibnamefont
				{Zou}},\ and\ \bibinfo {author} {\bibfnamefont {G.}~\bibnamefont {Guo}},\
		}\bibfield  {title} {\bibinfo {title} {Quantum illumination with
				photon-subtracted continuous-variable entanglement},\ }\href
		{https://doi.org/10.1103/PhysRevA.89.062309} {\bibfield  {journal} {\bibinfo
				{journal} {Phys. Rev. A}\ }\textbf {\bibinfo {volume} {89}},\ \bibinfo
			{pages} {062309} (\bibinfo {year} {2014})}\BibitemShut {NoStop}%
		\bibitem [{\citenamefont {Liu}\ \emph {et~al.}(2021)\citenamefont {Liu},
			\citenamefont {Tian}, \citenamefont {Liu}, \citenamefont {Dong},
			\citenamefont {Guo}, \citenamefont {He}, \citenamefont {Xu},\ and\
			\citenamefont {Li}}]{Imaging}%
		\BibitemOpen
		\bibfield  {author} {\bibinfo {author} {\bibfnamefont {D.}~\bibnamefont
				{Liu}}, \bibinfo {author} {\bibfnamefont {M.}~\bibnamefont {Tian}}, \bibinfo
			{author} {\bibfnamefont {S.}~\bibnamefont {Liu}}, \bibinfo {author}
			{\bibfnamefont {X.}~\bibnamefont {Dong}}, \bibinfo {author} {\bibfnamefont
				{J.}~\bibnamefont {Guo}}, \bibinfo {author} {\bibfnamefont {Q.}~\bibnamefont
				{He}}, \bibinfo {author} {\bibfnamefont {H.}~\bibnamefont {Xu}},\ and\
			\bibinfo {author} {\bibfnamefont {Z.}~\bibnamefont {Li}},\ }\bibfield
		{title} {\bibinfo {title} {Ghost imaging with non-gaussian quantum light},\
		}\href {https://doi.org/10.1103/PhysRevApplied.16.064037} {\bibfield
			{journal} {\bibinfo  {journal} {Phys. Rev. Appl.}\ }\textbf {\bibinfo
				{volume} {16}},\ \bibinfo {pages} {064037} (\bibinfo {year}
			{2021})}\BibitemShut {NoStop}%
	\end{thebibliography}
	
\end{document}